# High robustness quantum walk search algorithm with qudit Householder traversing coin, machine learning study

Hristo Tonchev[1,2*] and Petar Danev[2]

[1*]Institute of Solid State Physics, Bulgarian Academy of Sciences, 72 Tzarigradsko Chaussée, Sofia, 1784, Bulgaria.
[2]Institute for Nuclear Research and Nuclear Energy, Bulgarian Academy of Sciences, 72 Tzarigradsko Chaussée, Sofia, 1784, Bulgaria.

*Corresponding author(s). E-mail(s): htonchev@issp.bas.bg;

#### Abstract

In this work the quantum random walk search algorithm with walk coin constructed by generalized Householder reflection and phase multiplier has been studied. The coin register is one qudit with arbitrary dimension. Monte Carlo simulations, in combination with supervised machine learning, are used to find walk coins making the quantum algorithm more robust to deviations in the coin's parameters. By applying deep neural network we make prediction for the parameters of an optimal coin with arbitrary size and estimate the stability for such coin.

## 1 Introduction

Many examples of difference between the classical and quantum world can be given. Quantum Random Walk (QRW) [1] is one of them. Quantum interferences allow traversing graphs quadratically faster compared to classical walk [2]. This was initially tested on simple structures like line [3] and circle [3], later QRW is used to study more complex structures like square and hexagonal grids [4, 5], cylinder [6], torus [4], and hypercube [7]. Faster hitting time and ability to traverse arbitrary structures makes QRW a good base for variety of quantum algorithms like the one for finding triangles in graph [8], calculating Boolean formulas [9], quantum unsupervised machine learning [10], quantum neural networks [11], and the quantum random walk search algorithm (QRWS) [12]. The algorithm is used to search in different topologies like simplex and star like graphs [13], tree [14], square grid [4] and hypercube [15]. Due to diversity of tasks where QRW is used, variety of experimental implementations were considered like optical quantum computers [16], optical lattice [17], circuit QED [18], and ion traps [19].

Most quantum algorithms are constructed by using two state systems – qubits. However, using qudits instead of qubits has many advantages. Most obvious of them is exponential increasement of the databases with preserving the number of information carriers [20]. Other advantages of using qudits are that they give more robust to external noise quantum gates [21], more secure semi-quantum [22] and quantum [23, 24] cryptographic protocols, and increase in the effectiveness of variety of quantum algorithms. Examples for such algorithms are Shor's factoring [25], Grover's search [26], and quantum counting [27]. Qudits are also used in some quantum random walk based algorithms like Boolean formula evaluation [9] and quantum unsupervised machine learning [10].

There are two main methods for generating an arbitrary d-dimensional quantum gates used to construct algorithms with qudits - by Given's rotations [28] and by Householder reflections [29]. Decomposing quantum gates to Householder reflections is quadratically faster and can be used for qudits with arbitrary dimension [30]. It can be implemented effectively on various physical system like ion traps [31] and photonic quantum computer [32].

In the case of linear ion traps cooled to very low temperature and individual addressing of the ions, construction of qudit gates requires some additional conditions have to be met. For example, all Rabi frequencies of the ions interacting with the laser fields to have the same time dependence. Householder reflection also can be applied on qudits, when qudits are created by multipod system [26]. All qudit states are metastable states of the ion and transitions between levels are trough ancilla state. This technique requires precise control of the parameters of the laser field [31] (like pulse shape and detuning). Some techniques for reducing the sensitivity to such type of errors already exist like composite pulses [33], that can be useful in many applications. However, for other applications composite pulses alone will not be enough to reduce the error to appropriate level, in order for the algorithm to work properly.

In our previous work [34], we have studied discrete time QRWS algorithm with walk coin constructed by one generalized Householder reflection and phase multiplier. Such construction, with appropriately chosen parameters, leads to much more robust QRWS to inaccuracies in the phase multipliers. However, we used qubit coin, so we were restricted to coin sizes which are power of two – 2, 4, and 8.

In this paper we expand our previous work by studying QRWS algorithm with coin made by qudits. This reduces "distances" between coin sizes and gives us more data that allows us to study in more detail basic characteristics of the algorithm. We present quantitative description of the QRWS algorithm's stability with the proposed by us alternative construction of the walk coin. We also make predictions how well our methods will scale with increasing the coin size.

This paper is organized as follows: In section 2 the quantum random walk search algorithm and its modification that we propose in our previous work [34] have been described. First, in 2.1 a brief description of quantum random walk search algorithm and its quantum circuit is given. Next, in 2.2 we show the definition of the Householder reflection and its use in the construction of the walk coin. The advantage of using qudits for construction of the walk coin is explained in Sec. 3. The two subsections discuss the methods described in our previous work [34]. Here however, a qudit coin is used which allows us to gather much more data and obtain original results. In 3.1 our Monte Carlo method for studying QRWS algorithm is explained and few examples with qudit coin are given. Next, in 3.2 are shown functional dependences derived by MC simulation used to make QRWS algorithm more robust to inaccuracies in the parameters of the walk coin. New results for the behavior of those functions are presented for the first time here. Section 4 shows our main results in this paper. In 4.1 we analyze how the region of stability changes with increasing the coin size. Calculations of the robustness of the algorithm against the changes of its walk coin parameters are given in 4.2. The paper finishes with its conclusion 5.

# 2  Quantum random walk search with alternative walk coin

## 2.1  Quantum random walk search algorithm - quantum circuit

Quantum random walk search algorithm is quantum algorithm designed for searching in unsorted database with arbitrary topology. Because of the faster traversing of structures by the quantum random walk, QRWS is quadratically faster than its classical counterpart. Algorithm is probabilistic with probability to find solution depending of the node register size [12], topology of the searched structure [4, 15], modification of the algorithm [15], the number and position of the solutions [35, 36]. Quantum circuit of the QRWS algorithm is shown on Fig. 1.

The algorithm begins with preparing the initial states of node and edge registers in equal weight superposition. It continues by applying of algorithm's iteration fixed number of times. Probability to find solution is periodic function of the number of iterations. Each iteration consists of the following steps:

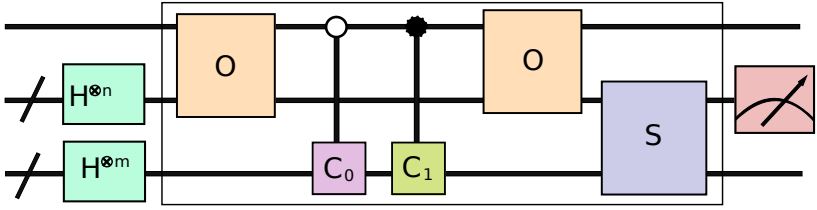

**Fig. 1** (Color online) Circuit of quantum random walk search algorithm.

applying the oracle operator $O$ that can recognize the solution, followed by marking $C_1$ and traversing $C_0$ coins that are applied depending on the states that oracle marks, next oracle should be applied second time, and Shift operator $S$ that define topology of searched structure should be used at the end of each iteration. The algorithm ends with measure of the node register. If measurement's result didn't give the searched element, algorithm is repeated.

Original QRWS on the hypercube uses Grover coin as traversing coin and have probability to find solution approximately $1/2 - O(1/2^m)$, where $m$ is the size of the edge register (and respectively node register has $n = 2^m$ states). The number of iterations $k$ of the QRWS algorithm for hypercube in case of one solution is obtained analytically when the Grover coin is used [15]:

$$k = \left[ \frac{\pi}{2} \sqrt{2^{m-1}} \right]$$

(1)

This formula can be used for coin consisting of qubits or one qudit with arbitrary dimension.

The maximum probability to find solution depends on the walk coin operator. In the next chapter we will briefly summarize the alternative construction of the walk coin by using Householder reflection and phase multiplier. More detailed description can be found in [34].

## 2.2　Walk coin by Householder reflection and additional phase multiplier

The traverse coin on undirected graph, can be constructed by using one generalized Householder reflection with phase $\phi$ and one phase multiplier $\zeta$ in the front of the coin, as is explained in [34]. The modified coin operator is:

$$C_0\,(\phi, \chi, \zeta) = e^{i\zeta}(I - (1 - e^{i\phi})|\chi\rangle\langle\chi|).$$

(2)

If the state used to build Householder reflection $|\chi\rangle$ is equal weight superposition, then the probability to go at each adjacent node is equal.

Here we will study the dependence of probability to find solution on the phases $\zeta$ and $\phi$. In our previous work we have shown that properly chosen relation between them will lead to more robust QRWS regardless of the node register's number of states. This will increase the probability in all experimental implementations where Householder reflection can be done efficiently.

Two simple examples for the coin operator $C_0$ with different values of angles are Grover coin when $\phi = \zeta = \pi$ (high probability) and identity operator when $\phi = \zeta = 0$ (there is no random walk). In those cases, the probability to find solution of the QRWS depends only on the number of states of node register $2^m$. However, in the general case the dependence is more complex:

$$p(\phi, \zeta, m) = \frac{1}{2} - \mathcal{O}\left(\frac{1}{2^m}\right) f(\phi, \zeta) \tag{3}$$

For searching in hypercube with dimension $m$, the walk coin should be made by qudits. If it consists of qubits - $m$ will be power of two, for qutrits – power of three, and so on. It is easy to see that the only thing that matter here is the total coin register size, not the way it is achieved – results of the simulations are the same for coin register consisting for 2 qutrits or one qudit with size 9. Further in this work we will consider the case when the coin register is built by only one qudit with arbitrary dimension. Node register of the algorithm will be built by using qubits.

# 3     QRWS with qudit coin constructed by Householder reflection

Contrary to the studies in our previous work, where the coin register consists of up to 3 qubits (coin size is power of two – 2, 4, 8), here we include cases with one qudit coin with dimension up to 11 states. Using qudits gives us more detailed information how the algorithm's parameters change with increase of the coin size.

## 3.1  Monte Carlo simulations of QRWS

We carried out series of Monte Carlo simulations of QRWS algorithm with qudit coin. Our numerical code takes random values for both angles $\phi, \zeta \in [0, 2\pi]$, and simulates the quantum algorithm with walk coin constructed by Householder reflection and phase multiplier with those values. The coin size is prior known, so we will consider it as constant ($m = const$). After each simulation of the algorithm the angles ($\phi, \zeta$) and the probability to find solution $p(\phi, \zeta)$ are stored.

Examples of Monte Carlo simulations for coins with dimension 5 and 9 are shown on Fig. 2(left) and Fig. 2(right) correspondingly. Darker color shows higher probability to find solution, and lighter color - smaller $p(\phi, \zeta)$. In case of coin size $m \geq 4$, there is a stripe with high probability to find solution. As $p(\phi, \zeta)$ increases the width of the stripe decreases. In Appx. A on Fig. A1, similar figures for coin size $m \in [2,10]$ are shown. For $m \geq 4$, they have the same behavior as the discussed above.

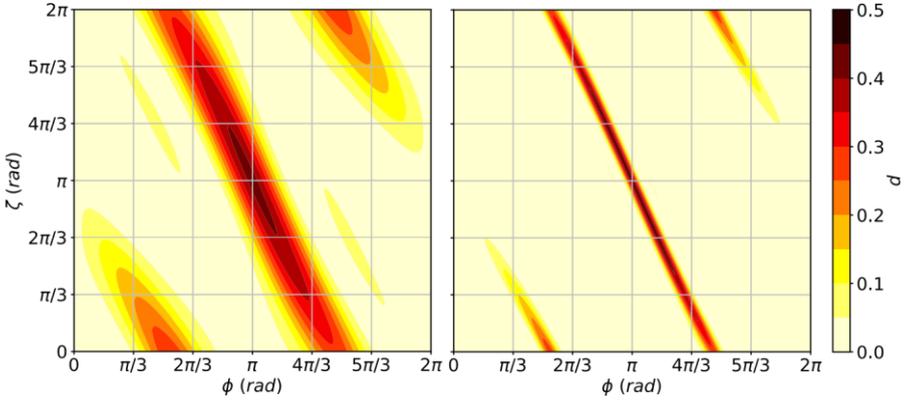

**Fig. 2** (Color online) Probability to find solution for coin sizes $m = 5$ (left) and $m = 9$ (right). The probability $p$ is plotted as a function of the angles $\phi$ and $\zeta$.

By using machine learning model trained on the Monte Carlo datapoints $(\phi, \zeta, p(\phi, \zeta))$ with coin size $m \in [2,10]$, we make a prediction for $p(\phi, \zeta)$, $m \geq 11$. The results for coin size 11 and 16 are plotted on the middle and right graphs of Fig. 3 respectively. They show that machine learning also confirms the observation from the last paragraph for reducing the width of the stripe of high $p(\phi, \zeta)$ with increasing the coin size. In order to evaluate the goodness of our ML model, in the case of QRWS with coin size $m = 11$, additional Monte Carlo simulations were done. The results are given on Fig. 3(left). If we compare it with the machine learning predictions Fig. 3(middle), except of slight blurring in the ML extrapolations explained in Appx. F, both figures show quite similar behavior.

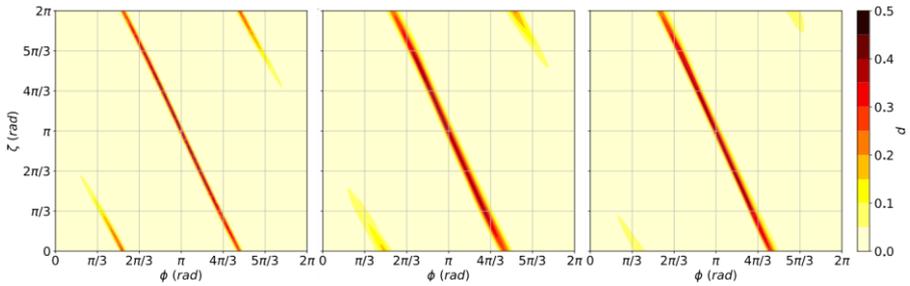

**Fig. 3** (Color online) Probability to find solution $p(\phi, \zeta, m)$ for coin sizes $m = 11$ (left and middle) and $m = 16$(right). The left image shows the Monte Carlo simulation results and the ones in the middle and on the right - machine learning predictions.

This observation points us to define a function $\zeta = \zeta(\phi)$, like in [34], and the probability to find solution becomes a curve $p(\phi)$ in the plane spanned by $\zeta$ and $\phi$:

$$p(\phi, \zeta, m) \rightarrow p(\phi, \zeta(\phi), m = const) \equiv p(\phi) \qquad (4)$$

## 3.2  Robustness of the coin for different functions $\zeta(\phi)$

Let $p_{max}$ is the maximum probability to find solution (for particular $m = const$) and $\phi_{max}$ is the value of $\phi$ where $p_{max}$ is achieved. Let the phase $\phi$ vary in the interval defined by $\Delta = (\phi_{max} - \varepsilon_-, \phi_{max} + \varepsilon_+)$, where $\varepsilon_-$ and $\varepsilon_+$ are different for different angle dependences $\zeta(\phi)$ and coin sizes. In this case the following equality holds:

$$p(\phi \in (\phi_{max} - \varepsilon_-, \phi_{max} + \varepsilon_+)) \cong p_{max} = p(\phi_{max}) \qquad (5)$$

Our goal is to find function $\zeta(\phi)$ that gives the largest possible interval $\Delta$. This will make the QRWS algorithm more robust to variation of the parameter $\phi$.

In our previous work [34] we have studied two types of functions. The linear functions:

$$\zeta = -2\phi + 3\pi \qquad (6)$$

$$\zeta = \pi \qquad (7)$$

were considered because the former (Eq. (6)) gives the best linear fit. The latter line (Eq. (7)) shows the case when parameter $\zeta$ is constant and this is a representation of the case when during the experiment no special relation between $\phi$ and $\zeta$ is preserved.

The second type of function we have considered is:

$$\zeta = -2\phi + 3\pi + \alpha \sin(2\phi), \phi \epsilon [0,2\pi]. \tag{8}$$

Such non-linear function gives better results in comparison to the linear functions shown above. However, here we introduce another parameter $\alpha$. The best value of $\alpha$ depends on the coin size, but relatively good initial choice is $\alpha = -1/(2\pi)$, as will be proven later in this paper. Such curve better approximates $p(\phi)$, however sinus function would be more difficult to implement in the experiments. The optimal values of the parameter $\alpha$ for each coin size $m - \alpha_{ML}(m)$, $2 \leq m \leq 16$ are given in Table 1. For coin size $2 \leq m \leq 11$, they are extracted from the numerical simulations' data and for $11 \leq m \leq 16$ (primed values) – from predictions of a deep network model, trained with the Monte Carlo datapoints for $2 \leq m \leq 10$. Detailed explanation of the DNN models used in the paper is given in Appendix F. In the case of coin with size 11, we give two relatively close values of $\alpha_{ML}$. They are derived independently, first from the MC simulations' data, and the second primed $\alpha_{ML}$ comes from machine learning predictions. The value $\alpha = -1/(2\pi) \simeq -0.159$ lies close to all $\alpha_{ML}(m)$, $m \geq 4$ which justifies its use as a benchmark throughout the paper.

| $m$ | 2 | 3 | 4 | 5 | 6 | 7 | 8 | 9 |
|---|---|---|---|---|---|---|---|---|
| $-\alpha_{ML}$ | 0.558 | 0.552 | 0.142 | 0.155 | 0.163 | 0.209 | 0.206 | 0.185 |
| $m$ | 10 | 11 | 11 | 12 | 13 | 14 | 15 | 16 |
| $-\alpha_{ML}$ | 0.168 | 0.150 | 0.170' | 0.179' | 0.180' | 0.203' | 0.225' | 0.197' |

**Table 1** Values of the parameter $\alpha_{ML}(m)$ in Eq. (8) for QRWS algorithm's walk coin size $2 \leq m \leq 16$. The primed values are derived from machine learning predictions, and the remaining come from Monte Carlo simulations.

Probability to find solution for the functions specified above in the case of coin size 5 and 9 is shown on Fig 4. Numerical simulations of equations (6) (red dot-dashed line), (7) (teal dashed line), and Eq. (8) with $\alpha = -1/(2\pi)$ (blue dotted line) were made, and the value $\alpha = \alpha_{ML}(m)$ is predicted by ML(solid green line).

In Appendix B on Fig. B2 are shown the probabilities $p(\phi, m)$, $m \in [2,11]$, for different functions $\zeta(\phi)$ (the same as in Fig. 4).

A comparison between the simulation results $p(\phi, m = 11)$ of QRWS algorithm and predictions of the trained neural network for the different relations $\zeta(\phi)$ is made to verify the reliability of the DNN extrapolations given in the paper. On Fig. 5(left) are drawn the lines from Monte Carlo simulations and on Fig. 5(middle) - are the predictions of the machine learning model. It can be seen that the curves on both

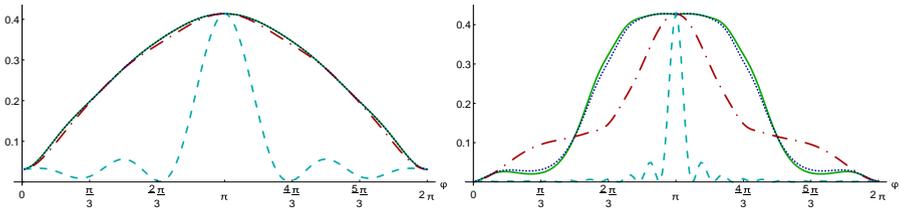

**Fig. 4** (Color online) The probability to find solution for coin sizes 5 and 9 in case of different functions $\zeta(\phi)$. The red dot-dashed line corresponds to Eq. (6), the blue dotted line - to Eq. (8) with $\alpha = -1/(2\pi)$, the teal dashed line - to Eq. (7), the solid green line - to Eq. (8) with $\alpha$ obtained by ML.

figures show the same behavior like, for example, the similar shape of the different lines (the solid green, dotted blue, and dot-dashed red corresponding to $\alpha = \alpha_{ML}$, $\alpha = -(2\pi)^{-1}, \alpha = 0$ in Eq. (8) respectively, and the dashed teal line - to Eq. (7)), although there are some asymmetries in the curves from the ML model. The lines on Fig. 5(right) represent predictions of the neural network for QRWS algorithm with walk coin size $m = 16$. By comparing both ML extrapolations for $m = 11$ (Fig. 5(middle)) and $m = 16$ (Fig. 5(right)), it is visible that all lines became steeper and there is an increase of the high probability $p(\phi, m)$ plateau with the increase of $m$, an observation supported by the MC simulations for lower coin register size (Fig. 4). Those results give us the premise to argue that the QRWS build with alternative walk coin proposed in [34] and thoroughly studied in this work, will give more robust implementation of the quantum algorithm for large coin register.

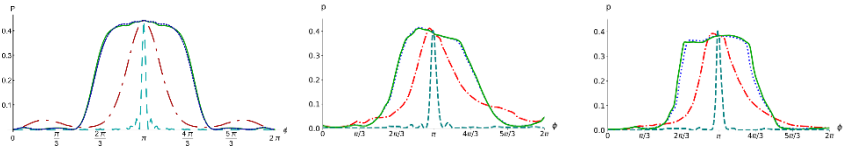

**Fig. 5** (Color online) The probability to find solution $p(\phi, m)$ for different functions $\zeta(\phi)$: the red dot-dashed line corresponds to Eq. (6), the blue dotted line - to Eq. (8) with $\alpha = -1/(2\pi)$, the teal dashed line - to Eq. (7), the green line - to Eq. (8) with $\alpha$ obtained by machine learning. The left picture shows results from Monte Carlo simulations for coin size $m = 11$, the middle one - ML predictions for $m = 11$, and the one on the right side - ML predictions for $m = 16$.

Based on our numerical results, we predict that with increasing the coin size, the curves for probability $p(\phi)$ given by equation (8) with $\alpha = -1/(2\pi)$ and $\alpha = \alpha_{ML}(m)$ will became indistinguishable. Those curves will become steeper and close to trapezoids.

The simulations of the algorithm with coin constructed by using relation (8) result in higher $p_{max}$ in comparison with the standard construction (with constant phase factor $\zeta = \pi$ as in Eq. (7)). In Table 3.2 are shown the maxima of probability $p(\phi, \zeta(\phi), m)$, for each of the functions $\zeta(\phi)$ (Eq. (6), Eq. (7), and Eq. (8) with $\alpha = -1/(2\pi)$ and $\alpha = \alpha_{ML}(m)$) and coin size $\phi \epsilon [0, 2\pi]$.

| Line \ Coin size | 4 | 5 | 6 | 7 | 8 | 9 | 10 | 11 |
|---|---|---|---|---|---|---|---|---|
| Eq. (7) | 0.3906 | 0.4137 | 0.4117 | 0.4022 | 0.4344 | 0.4272 | 0.4334 | 0.4414 |
| Eq. (8) & $\alpha = 0$ | 0.3906 | 0.4137 | 0.4117 | 0.4022 | 0.4344 | 0.4272 | 0.4334 | 0.4414 |
| Eq. (8) & $\alpha = (-2\pi)^{-1}$ | 0.3921 | 0.4137 | 0.4117 | 0.4082 | 0.4344 | 0.4279 | 0.4354 | 0.4414 |
| Eq. (8) & $\alpha = \alpha_{ML}$ | 0.3921 | 0.4137 | 0.4117 | 0.4093 | 0.4344 | 0.4277 | 0.4344 | 0.4414 |

**Table 2** Maxima of probability to find solution by coin using relation (7) is shown on the first row. Analogically, $p$ corresponding to Eq. (8), with $\alpha = 0$ is shown on the second row, and with $\alpha = -1/(2\pi)$ - on the third row. The fourth row corresponds to equation (8) with $\alpha$ obtained by ML.

As the coin size increases the probability to find solution becomes higher. In the case of Eq. (6) and Eq. (7) for coin size greater than 3 the maximum is always at the point with coordinates $\phi = \pi$ and $\zeta = \pi$ (corresponding to Grover's coin). The lines given by Eqs. (6), (7), and (8) have axial symmetry around line $\phi = \pi$. However, the $p_{max}$ is not always at this point. The equation $\zeta = -2\phi + 3\pi + (-1/(2\pi))sin(2\phi)$ gives higher $p(\phi)$ in comparison to the Grover coin for coin sizes $m = 4,7,9,10$. For example, in the case of $m = 7$ the maxima are at the points $\{2.7925 \pm 0.0349, 3.49066 \pm 0.0349\}$ and the increasement is significant.

# 4   Numerical results

## 4.1   Region of stability for different coin sizes

During the experiments there are time variations of the experimental setup's parameters, like in the frequency and the shape of the laser pulses controlling ions in the trap, etc. In order to investigate the robustness of the implementation of QRWS algorithm we will study the half length $\varepsilon = (\varepsilon_- + \varepsilon_+)/2$ of the interval $\Delta$.

Here we show our results for the width $\varepsilon$ of each of the functions (6), (7), and Eq. (8) with $\alpha = -1/(2\pi)$ and $\alpha = \alpha_{ML}(m)$ , where $m \in [2,16]$ . The solid lines correspond to results from MC simulation, and dashed lines - to prognosis given by machine learning. In Appx. F, an explanation of ML methods used and their drawbacks are given. Different colors and markers correspond to different dependence between phases in the walk coin. The red curve with 4-pointed star marker corresponds to Eq. (6), the teal with 3-pointed star marker – to Eq. (7), and the blue with 5-pointed star marker and the green with 2-pointed star marker – to Eq. (8) with $\alpha = -1/(2\pi)$ and $\alpha = \alpha_{ML}(m)$ correspondingly. We will make the remark that MC points for $m = 11$ were not enough to obtain good ML estimation of $\alpha$. This is the reason why we used $\alpha_{ML}(11)$ predicted with neural network model and used it in QRWS algorithm simulation. This approach gives significantly better estimation for $\varepsilon$ in the case $m = 11$. As an example, on Fig. 6 is shown the value of $\varepsilon$ in the cases when $p(\phi \in (\phi_{max} - \varepsilon_-, \phi_{max} + \varepsilon_+)) \geq 0.9 \times p_{max}$ and analogically for $0.7 \times p_{max}$:

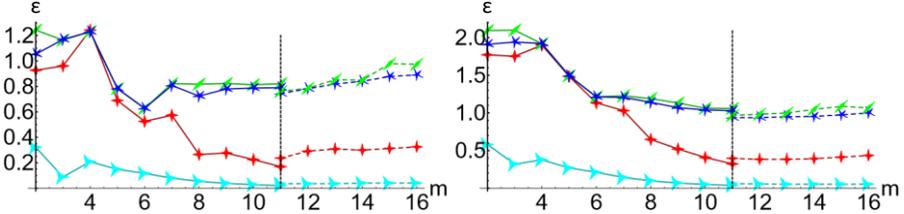

**Fig. 6** (Color online) Change of the area width $\varepsilon$, having probability to find solution equal to percentage of the maximum probability $p_{max}$, with increase of the coin size $m$. The left figure shows the case for $90\% \times p_{max}$ and the right - for $70\% \times p_{max}$. The red curve with 4-pointed star marker corresponds to Eq. (6), the teal with 3-pointed star marker to Eq. (7), and the blue with 5-pointed star marker and the green with 2-pointed star marker to Eq. (8) with $\alpha = -1/(2\pi)$ and $\alpha = \alpha_{ML}(m)$ correspondingly. Solid lines correspond to values obtained by MC simulations, dashed to - results prognosed by ML.

The value of $\varepsilon$, depends on the coin size and on the function used to construct the coin. Monte Carlo simulation of QRWS by using coins with size $m \leq 3$ have completely different behavior compared with the larger coins as can be seen in Appendix A on Fig. A1. For those cases the Grover coin gives very low probability to find solution. This is the reason to exclude MC simulations of QRWS with coin size $m \leq 3$.

In case of equation (7) the initial length $\varepsilon$ is low and continue to decrease with increasing coin size. The curve from Eq. (6) for small coins have high width $\varepsilon$, however it decreases fast with increasing the coin size. For large coin we predict that they will have similar behavior and Eq. (7) will act like Eq. (6) for smaller coin. So, the linear approximation is not good enough.

Equation (8) with $\alpha = 0$ coincides with Eq. (6). However, other values of $\alpha$ are much more promising. The probability $p$ for coin constructed with functional dependence $\zeta(\phi)$, given by Eq. (8), didn't seem to decrease if coin size increases. The curve corresponding to $\alpha = \alpha_{ML}(m)$ obtained by ML gives a slightly larger $\varepsilon$ and shows slightly better behavior with increase of the coin size compared to $\alpha = -1/(2\pi)$. The value of $\varepsilon$ deceases between 3 and 6, and remains approximately the same thereafter ($\varepsilon \simeq \pi/4$) as can be seen on FIG. 6. The behavior of $\varepsilon$ is different when we evaluate it for different percentage of the maximum probability. Two examples for $0.9 \times p_{max}$ and $0.7 \times p_{max}$ are shown on FIG. 6.

Based on Monte Carlo simulations we predict that $\varepsilon$, when $p(\phi \in (\phi_{max} - \varepsilon_-, \phi_{max} + \varepsilon_+)) \simeq 0.9 \times p_{max}$, and for coin constructed according to equations Eq. (7) and Eq. (6), will decrease with increasing the coin size and for large coin $\varepsilon \to 0$. We also predict that $\varepsilon$ for Eq. (8) with $\alpha = \alpha_{ML}$ will remain constant with increasing the coin size. The interval $\varepsilon$ for Eq. (8) with $\alpha = (-1/(2\pi))$ will became closer to $\varepsilon$ corresponding to $\alpha = \alpha_{ML}$ and for large coins $\alpha_{ML} \rightarrow (-1/(2\pi))$ . For lower percentage of the maximum height (for example $0.7 \times p_{max}$), $\varepsilon$ for all those functions will decrease slowly with increasing the coin size and for large coin they will converge to a fixed value. The observations given above are supported by ML simulations, when we consider some drawbacks of our ML models explained in the Appendix F.

The analysis above shows that the half-width $\epsilon$, did not decrease with increasing the coin register size if Eq. (8) with optimal parameters is used. This indicates that such modification of QRWS algorithm retains its high robustness even for large coin size.

## 4.2  Analysis of the the algorithm's robustness

The proposed in the work optimization of the quantum random walk search algorithm is based on walk coin parameterized by two phases (see Eq. (2)). We have showed in Sec. 3.2 that, if an optimal relation between them is maintained, the algorithm becomes more robust. Here we will prove this statement by numerically analyzing the stability of probability $p$ to uncertainties in the walk coin parameters $\phi$ and $\alpha = \alpha(\phi, \zeta)$. The latter expression is derived by solving Eq. (8) for $\alpha$.

On Fig. 7 simulations of the probability to find solutions $p(\phi, \alpha, m)$ are shown as a function of angle $\phi$, coin size $m = 5$(left), 9(right), and parameter $\alpha$ of Eq. (8). For walk coin with size $m \geq 4$ there is a central plateau with high values of the probability $p(\phi, \alpha, m)$. The lightest color area corresponds to $p > 0.95 \times p_{max}(m)$, the second contour is at $p = 0.9 \times p_{max}(m)$, etc. The values of $p_{max}(m)$ are given in Table 2. The plateau is relatively wide not only in the horizontal axis (along $\phi$), but in $\alpha$ direction too. And this behavior remains the same for all of the simulated QRWS schemes with walk coin size up to 10, as could be verified from the figures illustrating the function $p(\phi, \alpha, m), 2 \leq m \leq 10$ in Appendix D. The above observation shows that the proposed in paper construction of the coin leads to stable quantum algorithm against small deviations in both $\phi$, and the introduced by our scheme, parameter $\alpha$.

The horizontal lines on both pictures of Fig. 7 mark specific values for the angle $\alpha$. The dashed gray, dash-dotted blue, and the solid green lines on Fig. 7 correspond to $\alpha = 0$, $\alpha = -1/(2\pi)$, and $\alpha = \alpha_{ML}(m)$ (given in Table 1) respectively. The latter two lies close one to the other and both lie in regions with high probability $p$ for all of the simulated QRWS algorithms with walk coin size $m \geq 4$. This shows that the value $\alpha = -1/(2\pi)$ is relatively good and could be used in practice, even if the optimal value of $\alpha$ is not known precisely for quantum algorithms with even larger registers.

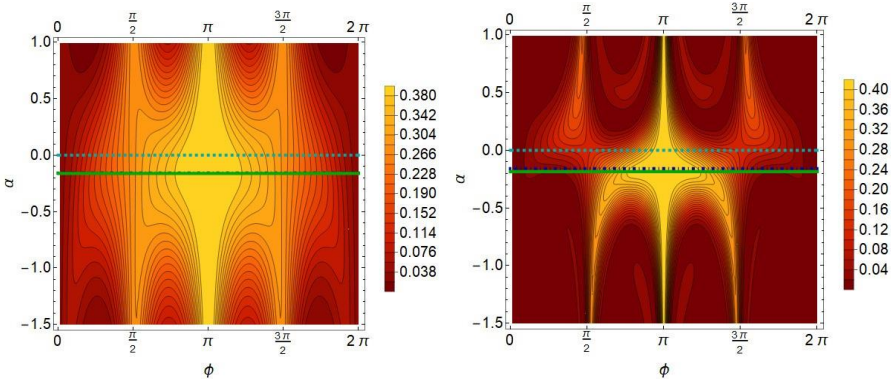

**Fig. 7** (Color online) Probability of QRWS algorithm to find solution $p(\phi, \alpha, m)$ as a function of $\alpha$ and $\phi$ for walk coin with size $m = 5$ (left) and $m = 9$ (right). The horizontal lines represent simulation of QRWS algorithm with $\alpha = 0$ - dashed gray line, $\alpha = -1/(2\pi)$ – dash-dotted blue line, and the solid green line corresponds to $\alpha_{ML}(m)$ given in Table 1.

In order to give quantitative proof for the increased robustness of the QRWS algorithm, we make estimation of the root-mean-square (r.m.s) deviation of the probability $p(\phi, \alpha, m)$ for a fixed uncertainty in the parameters $\alpha$ and $\phi$ – $\sigma_\alpha$ and $\sigma_\phi$. We do not assume any correlation between these parameters and calculate the quantity:

$$\sigma_p^{ij} = \frac{1}{p^{ij}} \sqrt{\left(\frac{\partial p^{ij}}{\partial \phi^i}\right)^2 \sigma_\phi^2 (\phi^i - \pi)^2 + \left(\frac{\partial p^{ij}}{\partial \alpha^j}\right)^2 \sigma_\alpha^2 (\alpha^j - \alpha_{ML}(m))^2} \qquad (9)$$

for each discrete point $\phi_i$ and $\alpha_j$ in the plane $(\phi, \alpha)$, where $0 \geq \phi \geq 2\pi$, $-1.5 \geq \alpha \geq 1$ with $i = 1, \ldots, 180$ and $j = 1, \ldots, 250$. We center the data around the points $(\pi, \alpha_m)$ which lie in the middle of the region giving the most stable implementation of the quantum algorithm. Here $\alpha_{ML}(m)$, $2 \leq m \leq 10$ is the optimal value of the parameter $\alpha$ in Eq. (8) obtained by fitting the expression to the numerical datapoints for QRWS with walk coin size $m$. Their values are given in Table 1.

The results for walk coin with size 5 (Fig. 8(left)) and 9 (Fig. 8(right)) are graphically represented in logarithmic scale. The computations are done with $\sigma_\alpha = 0.1$ and $\sigma_\phi = 0.1$. As can be seen from the figures, there is relatively large central area where the r.m.s. deviation of the probability $\sigma_p$ is less than 0.01, reaching levels $\sigma_p \leq 10^{-4}$ in the innermost dark region. This behavior is the same for quantum algorithm using walk coin with five and nine states. The corresponding images on Fig. E6 in Appendix E for coin with size $2 \leq m \leq 10$ show similar characteristics. The above analysis confirms that the alternative walk coin, constructed by Householder reflection and additional phase, improves QRWS algorithm, not only to obtain high probability $p(\phi, \alpha, m)$ for large interval of $\phi$, but also to have extremely high robustness to changes in both parameters of the coin.

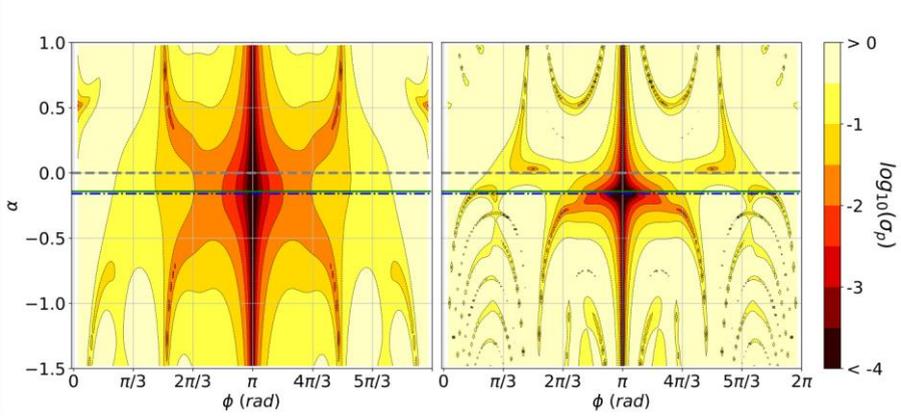

**Fig. 8** (Color online) R.m.s. deviation of the probability $p(\phi, \alpha, m)$, given with Eq. (9), for walk coin with size $m = 5$ (left) and $m = 9$ (right). The dark central area represents high robustness of the quantum algorithm for small deviations of the algorithm's parameters. The dashed gray, dash-dotted blue, and the solid green lines correspond to $\alpha = 0$, $\alpha = -1/(2\pi)$, and $\alpha = \alpha_{ML}$.

It is useful to study how the robustness of the proposed construction of the quantum random walk search algorithm compares to the stability of QRWS when the walk coin is built according to Eq. (7) - corresponding to the case when the parameter $\zeta$ is not controlled (if $\phi = \pi$ the Grover coin is obtained). We have calculated the r.m.s. of the probability for the latter case:

$$\sigma_{p'}^i = \frac{1}{p'^i} \left| \frac{\partial p'^i}{\partial \phi^i} \sigma_\phi(\phi^i - \pi) \right| \tag{10}$$

Here $\sigma_\phi = 0.1$ and $p'(\phi, m)$ is the probability to find solution corresponding to the curves $\zeta = \pi$ given by Eq. (7) (depicted with the dashed teal lines on both pictures of Fig. 4).

On Fig. 9 are plotted the ratios $\sigma_p/\sigma_{p'}$ in logarithmic scale, where the division is made for every value of $\sigma_j$ (i.e., $\sigma_p^{i\,j}/\sigma_{p'}^i$, where j = const). On the left and right sides of the picture are shown the results for coin size $m = 5$ and $m = 9$ respectively. We will make remarks on two important points. First, the alternative QRWS shows stability $\geq \mathcal{O}(10^2)$ in comparison to the most commonly used walk coin in the most important central region. Outside this region the probability of the standard QRWS algorithm $p(\phi, \zeta = \pi, m)$ is very low and comparison of the robustness is not so essential. Second, by carefully comparing the two images for coin size five and nine on Fig. 9 (all the pictures for $2 \geq m \geq 10$ from Fig. E7 in Appendix E show the same behaviour), it can be seen that the relative stability improves with the increase of the coin size.

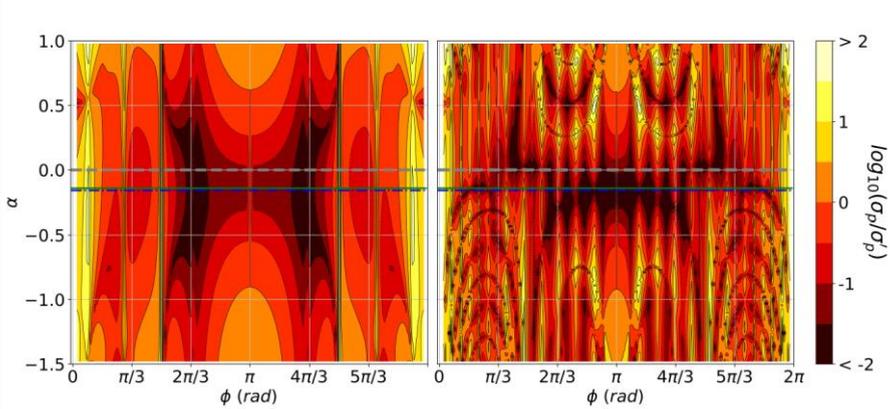

**Fig. 9** (Color online) Relative stability of the quantum random walk search algorithm with walk coin size $m = 5$ (left) and $m = 9$ (right), given by the ratio of the root mean square uncertainty with (Eq. (9)) and without (Eq. (10)) the proposed optimization of the coin. The dark central area represents high relative robustness. The dashed gray, dash-dotted blue, and the solid green lines correspond to $\alpha = 0$, $\alpha = -1/(2\pi)$, and $\alpha = \alpha_{ML}(m)$.

The latter is even more evident on Fig. 10, where the innermost central region of $\sigma_p/\sigma_{p'}$ for coin size $m = 5$ (left) and $m = 9$ (right) is presented (all studied cases are shown on Fig. E8 in Appx. E). The range in the parameter $\phi$ is chosen in such a way that it corresponds to the high probability central peak of $p(\phi, \zeta = \pi, m)$ given by Eq. (7).

For practical realizations of QRWS, the registers should have large number of qubits(qudits). Thus, the expected improvement of the relative stability of the algorithm for large walk coin register, would lead to more robust experimental implementation of the quantum random walk search algorithm. We note that the vertical stripes in light tones close to the central region on Fig. 9 (right) correspond to low $p'(\phi, m)$, but relatively high $p(\phi, \alpha, m)$, thus these areas could be ignored when considering the relative stability, and they do not diminish the validity of the above statements.

In this section we have studied in more detail the stability of the quantum random walk algorithm constructed with an alternative walk coin. The results show the existence of large area around the optimal coin parameter values with very small r.m.s. deviation of the probability $p(\phi, \alpha, m)$ to uncertainties in both angles $\phi$ and $\alpha$. When we compare the stability of the proposed QRWS algorithm to the one with the commonly used walk coin, a slight increase in the relative robustness is observed for higher coin register size.

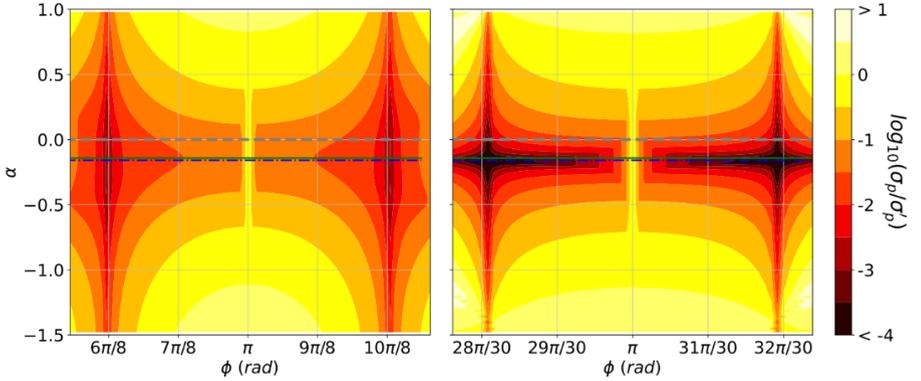

**Fig. 10** (Color online) Relative stability of the quantum random walk search algorithm with walk coin size $m = 5$ (left) and $m = 9$ (right), given by the ratio of the root-mean-square uncertainty with and without the proposed optimization of the coin. The chosen ranges of the parameter $\phi$ correspond to the high probability central peaks of $p(\phi, \zeta = \pi, m)$ (Eq. (7)). The dashed gray, dash-dotted blue, and the solid green lines are obtained from (Eq. (8)) with $\alpha = 0$, $\alpha = -1/(2\pi)$, and $\alpha = \alpha_{ML}$.

# 5 Conclusion

In this paper we have studied an alternative walk coin construction in the quantum random walk search algorithm by Householder reflection and an additional phase multiplier. In contrast to our previous work, where we have considered only coin build with qubits, here the more general qudit walk coin is studied. Monte Carlo simulations of QRWS algorithm for different size of the coin register were conducted. We have shown that, if a proper relation between the coin parameters is maintained, the quantum algorithm become much more robust, and this property remains true for every size of the quantum algorithm registers studied in the paper. We have investigated the behavior of practically important quantities like the width of the area, giving high probability to find solution and the maximum of this probability, for different coin parameters and relations between them. Optimal functional dependencies and parameter values are derived for walk coin with size up to one qudit with eleven states. Even more, by using machine learning methods, we have made extrapolations of the mentioned quantities for larger size of the coin. From analysis of the Monte Carlo data and the machine learning predictions, we show that the quantum random walk search algorithm with the proposed alternative walk coin, demonstrates high robustness to deviations of its parameters. We have calculated the stability of the proposed alternative construction of the quantum random walk search algorithm and the numerical results show that there exists wide area in the space spanned by the walk coin's parameters with extremely low root mean square uncertainty of the probability of QRWS algorithm to find solution. We have also studied the relative uncertainty of our construction of the QRWS algorithm to the uncertainty of QRWS algorithm with a standard walk coin and show that the former is more robust to coin parameter's deviations and that the relative stability increases for larger coin register.

# Acknowledgments

The work on this paper was supported by the Bulgarian National Science Fund under Grant KP-06-M48/2 / 26.11.2020.

# Appendix A     Monte Carlo simulations for different coin size

On Fig. A1 are shown Monte Carlo simulations' results of probability to find solution $p(\phi, \zeta, m)$ of Quantum random walk search algorithm with coin constructed by Householder reflection and additional phase multiplier. Simulation parameters are both phases $\phi$ and $\zeta$. Different pictures correspond to different size of the coin $m$. The images on the first row represent QRWS with coin size $m = 2,3,4$, on the second – with $m = 5,6,7$ and on the third – with $m = 8,9,10$.

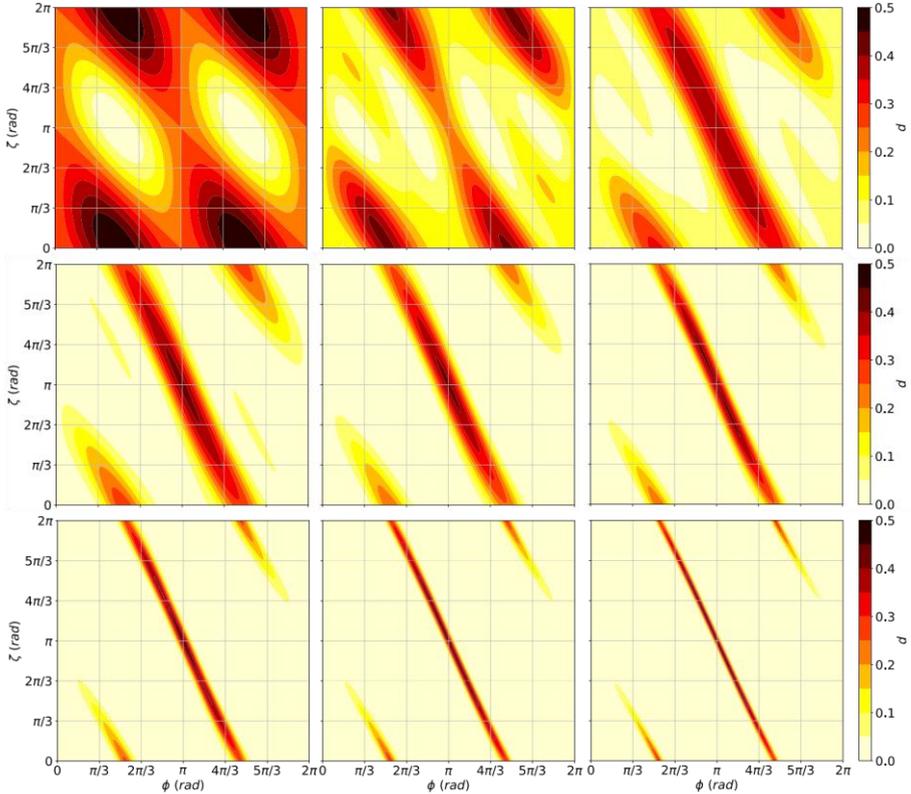

**Fig. A1** (Color online) Monte Carlo simulation of the probability to find solution $p$ of QRWS algorithm with coin constructed by Generalized Householder reflection and phase multiplier. Coordinate axes correspond to the phases $\phi$ and $\zeta$. On the first row are put pictures for coin size $m = 2,3,4$ (from left to right), on the second – for $m = 5,6,7$, and on the third – for $m = 8,9,10$. Higher probability is shown with darker color.

For coin size 2 areas with high probability to find solutions are chess like arranged rhomboids. For coin size 3 those rhomboids begin to merge in the direction of the line $\zeta = -2\phi + 3\pi$. At coin size 4, the merged rhomboids became parallel stripes. From this point the stripes' width begins to decrease with increasing the size of the coin. The results are consistent with the conclusions in our previous paper [34].

## Appendix B     Curves for different qudit size

On Fig. B2 is shown the probability to find solution $p(\phi, \zeta, m)$ for different relations between $\zeta$ and $\phi$. The red dot-dashed line corresponds to Eq. (6) and the teal dashed line - to Eq. (7). The blue dotted and the solid green lines show Eq. (8) with $\alpha = -1/(2\pi)$ and $\alpha = \alpha_{ML}(m)$ correspondingly. On the first row are given images for coin sizes 2, 3, and 4, on the second - 5, 6, and 7, on the third - 8, 9, and 10.

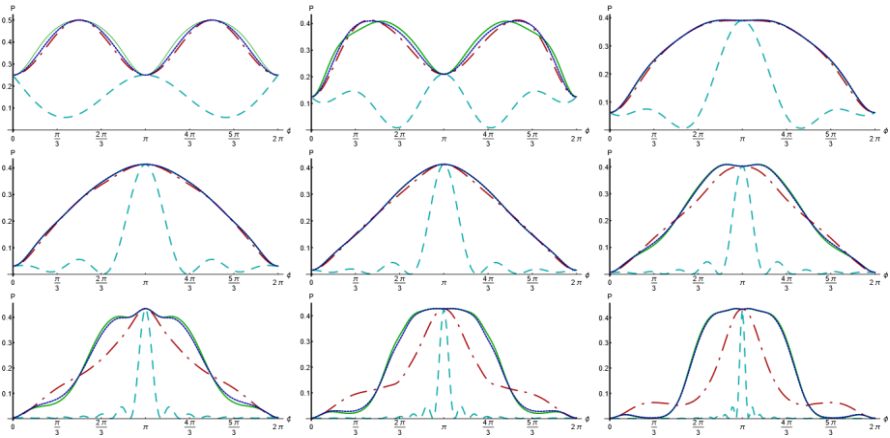

**Fig. B2** (Color online) Probability to find solution $p(\phi, m)$ for different coin size $2 \leq m \leq 10$. On the first row are positioned pictures for coin size $m = 2,3,4$ (from left to right), on the second – for $m = 5,6,7$, and on the third – for $m = 8,9,10$. Different curves correspond to different curve. The red dot-dashed line corresponds to Eq. (6), the blue dotted line - to Eq. (8) with $\alpha = -1/(2\pi)$, the teal dashed line - to Eq. (7), the green line - to Eq. (8) with $\alpha$ obtained by ML.

## Appendix C     Region of stability for different coin sizes – practical consideration

In section 4.1 the interval where probability to find solution exceeds certain percentage of its maximum value has been studied. More practical point of view is to investigate the interval $\varepsilon$ where probability $p$ is greater than fixed value $p > p_{LocM}(m)$. This can be used by the experimentalists, when their goal is to reach certain probability to find solution after each run of the algorithm. Results of simulations for $p \geq 0.37$ and for $p \geq 0.31$ are shown on the left and the right sides of

Fig. C3. Different colors correspond to different dependence between the phases in the walk coin. The red curve with 4-pointed star marker corresponds to Eq. (6), the teal with 3-pointed star marker – to Eq. (7), and the blue with 5-pointed star marker and the green with 2-pointed star marker correspond to Eq. (8) with $\alpha = -1/(2\pi)$ and $\alpha = \alpha_{ML}(m)$ correspondingly.

In case of coins obtained by Eq. (8) with $\alpha = \alpha_{ML}(m)$ and $\alpha = -1/(2\pi)$ we have better results. On Fig. C3(left) $\varepsilon$ for all curves decreases until reaching size 6, then it begins to slowly increase until reaching coin size 11. However, more information is needed to say with certainty how they will behave for larger coin sizes. On the right figure, $\varepsilon$ decreases fast until it reaches coin size 6 and then begins to slowly decrease.

The maximal probability to find solution increases with the coin size. So, a fixed value of $p$, with the increasement of the coin size, became smaller percentage of $p_{max}$. That results in different behavior of the curves compared to the ones on Fig. 6. The lines on Fig. C3 given by Eq. (8) with $\alpha = \alpha_{ML}(m)$ and $\alpha = -1/(2\pi)$ didn't converge to a fixed value and they continue to increase. From the above, we make conclusion that the robustness of the proposed algorithm modification increases if greater probability than fixed value is desired.

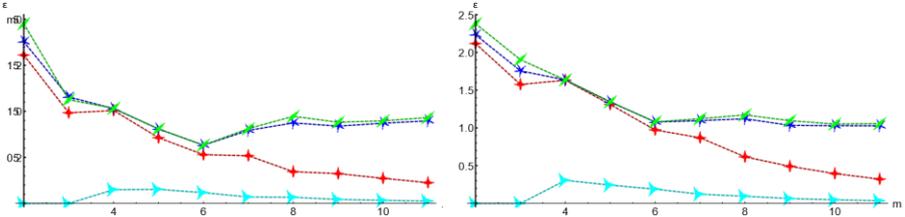

**Fig. C3** (Color online) Change of the width of the area with probability to find solution greater than 0.37 and greater than 0.31. The red curve (4-pointed star marker) corresponds to Eq. (6), the teal (3-pointed star marker) to Eq. (7) and the blue (5-pointed star marker) and the green (2-pointed star marker) to Eq. (8) with $\alpha = -1/(2\pi)$ and $\alpha = \alpha_{ML}(m)$ correspondingly.

## Appendix D     Area of stability of QRWS as function of $\alpha$ and $\phi$

On Fig. D4 is shown the probability to find solution $p(\phi, \alpha, m)$, when coin is constructed by generalized Householder reflection and phase multiplier (Eq. 2), and there is a relation between phases as in Eq. 8. The walk coin depends on phase $\phi$, parameter $\alpha$, and size of the coin $2 \leq m \leq 10$. Lighter colors correspond to higher $p$. The color scale counturs show probability between 5% and 95% of $p_{max}$.

More important case from practical point of view, analogically to Appendix C, is when we search for areas with probability $p$ greater than fixed value. The pictures on Fig. D5 show probability $p$ as function of $\phi$ and $\alpha$. However, contrary to D4, here fixed values of $p$ (between 0.037 and 0.37) are taken for the color scale contours.

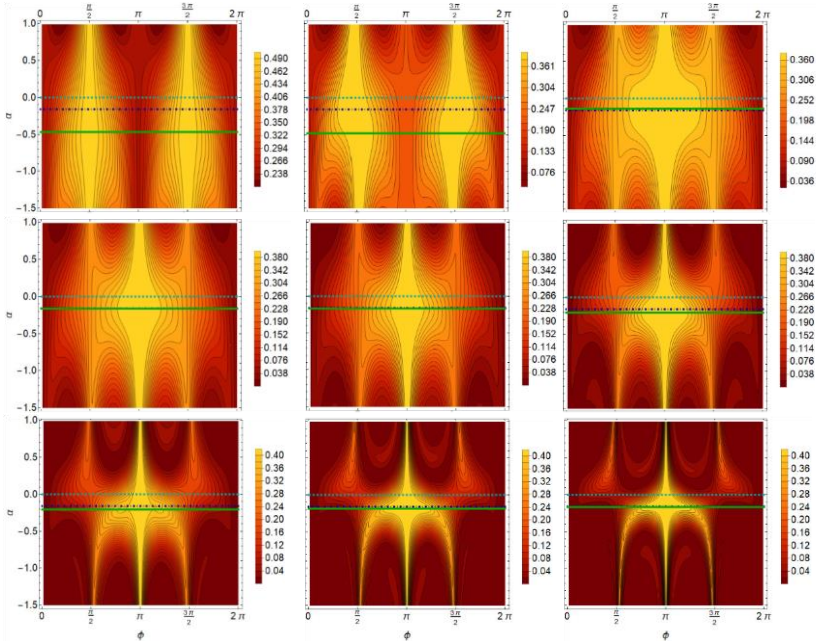

**Fig. D4** (Color online) Probability to find solution $p$ as function of $\phi$ and parameter $\alpha$ in the non-linear equation (8). The solid green line corresponds to $\alpha = \alpha_{ML}(m)$, the dashdoted blue – to $\alpha = -1/(2\pi)$ and dashed teal – to $\alpha = 0$. The lighter colors show higher probability and darker – lower. The scale is between 0 and $p_{max}$, and the contours are at intervals $5\% \times p_{max}$. The probabilities larger than $95\% \times p_{max}$ are depicted with the lightest color. Each picture corresponds to different value of $m \in [2,10]$. On first row from left to right are given the pictures for $m = 2,3,4$, on the second – $m = 5,6,7$, and on the last – $m = 8,9,10$.

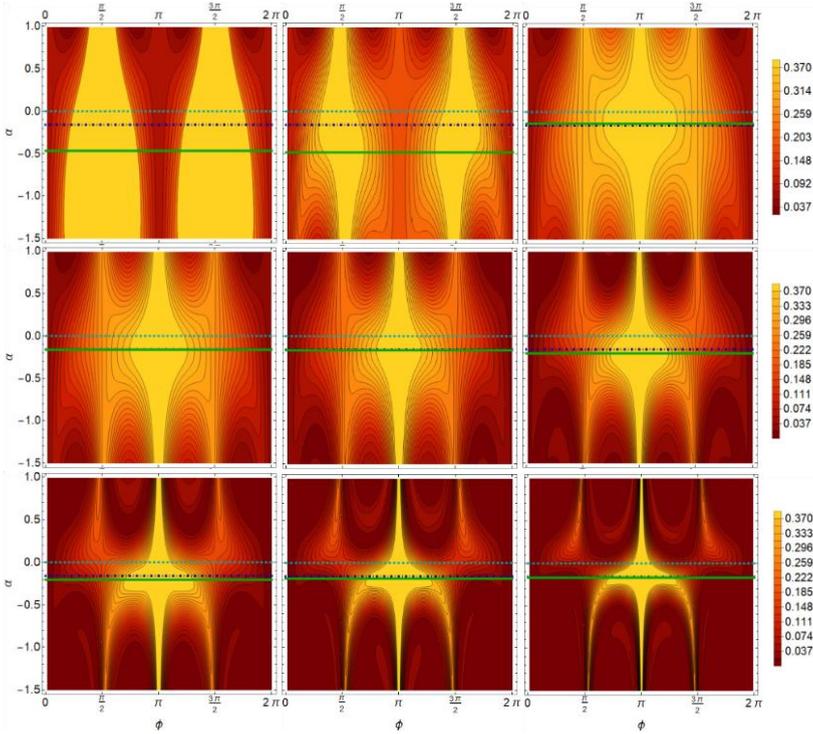

**Fig. D5** (Color online) Probability to find solution $p$ as function of $\phi$ and parameter $\alpha$ in the non-linear equation ([8]). The solid green line corresponds to $\alpha = \alpha_{ML}$ ($m$), the dashdoted blue – to $\alpha = -1/(2\pi)$ and dashed teal – to $\alpha = 0$. The lighter colors show higher probability and darker – lower. The scale is from 0 to $p_{max}$, and the ticks are between 0.037 and 0.37. Each picture corresponds to different value of $m \in$ [2,10]. On first row from left to right are given the pictures for $m = 2,3,4$, on the second – $m = 5,6,7$, and on the last – $m$=8,9,10.

# Appendix E    Robustness of the modified quantum random walk algorithm for different size of the walk coin

In Sec. 4.2 the stability of our proposed modification of the quantum random walk search algorithm was investigated. A quantitative description of the root-mean-square deviation of the probability to find solution and the relative improvement in the stability have been graphically presented in the cases of walk coin size 5 and 9.

Here, on Fig. E6 we give a complete set of results for the robustness of the QRWS algorithm, as defined in Eq. (9), for coin with size $2 \leq m \leq 10$. It can be seen that, in all of the pictures, there is wide central area (darker colors) where the relative probability $\sigma_p$ is extremely low for small changes in the parameters $\phi$ and $\alpha$.

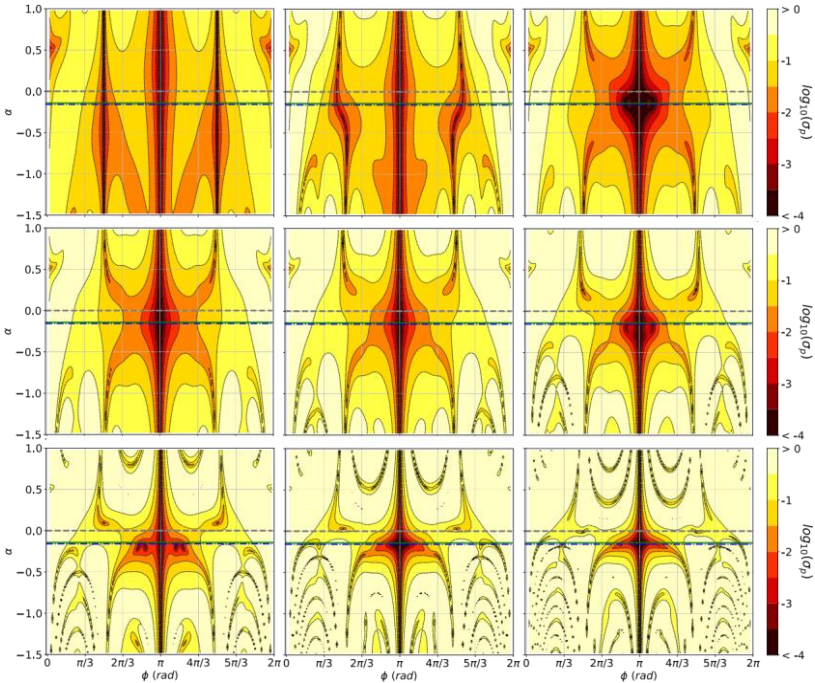

**Fig. E6** (Color online) Root-mean-square deviation of the probability $p(\phi, \alpha, m)$, given with Eq. (9), for walk coin with size $m = 2, 3, 4$ (first row, from left to right), $m = 5, 6, 7$ (second row), and $m = 8, 9, 10$ (third row). The dark central area represents high stability region of the quantum algorithm for small deviations of the algorithm's parameters $\phi$ and $\alpha$. The dashed gray, dash-dotted blue, and the solid green lines correspond to $\alpha = 0$, $\alpha = -1/(2\pi)$, and $\alpha = \alpha_{ML}$.

Analogously, on Fig. E7 are presented the numerical calculations' results for the ratio of the relative stability of the QRWS algorithm proposed by us in [34], to the one with standard construction of the walk coin. By increasing the coin size from $m = 2$ to $m = 10$, an increase of the central dark area can be observed. This indicates that with increase of the coin size, the algorithm's relative stability continues to improve.

The innermost central region of $\sigma_p/\sigma_{p'}$ for each studied coin size m is presented on Fig. E8. The range in the parameter $\phi$ is chosen in such a way that it corresponds to the high probability central peak of $p(\phi, \zeta = \pi, m)$ given by Eq. (7). On Fig. E8 the improvement in the relative robustness with the increase of the walk coin size is even more evident.

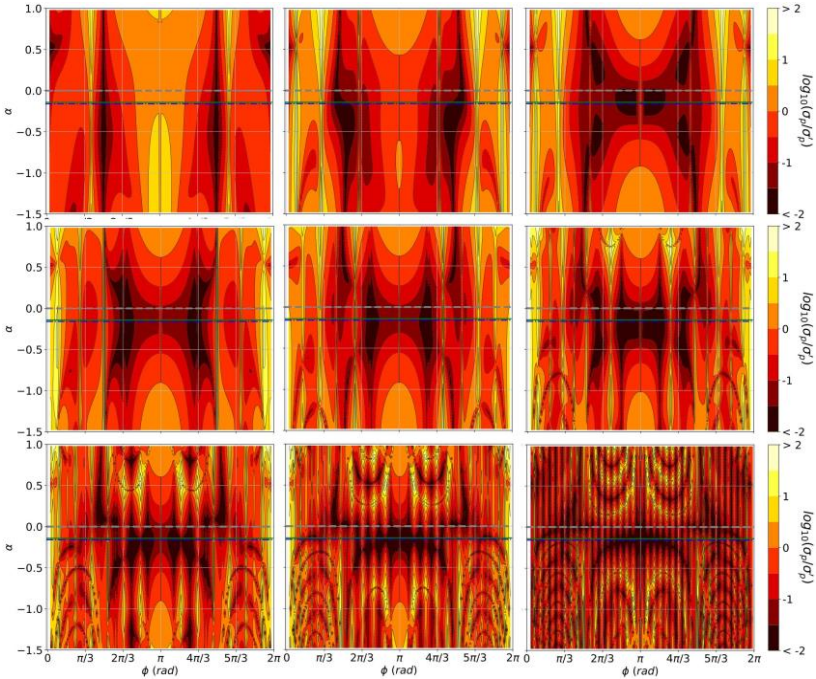

**Fig. E7** (Color online) Relative stability of the quantum random walk search algorithm with walk coin size $m = 2,3,4$ (first row, from left to right), $m = 5,6,7$ (second row), and $m = 8,9,10$ (third row), given by the ratio of the root-mean-square uncertainty with and without the proposed optimization of the coin. The dark central area represents high relative robustness. The dashed gray, dash-dotted blue, and the solid green lines correspond to $\alpha = 0$, $\alpha = -1/(2\pi)$, and $\alpha = \alpha_{ML}$.

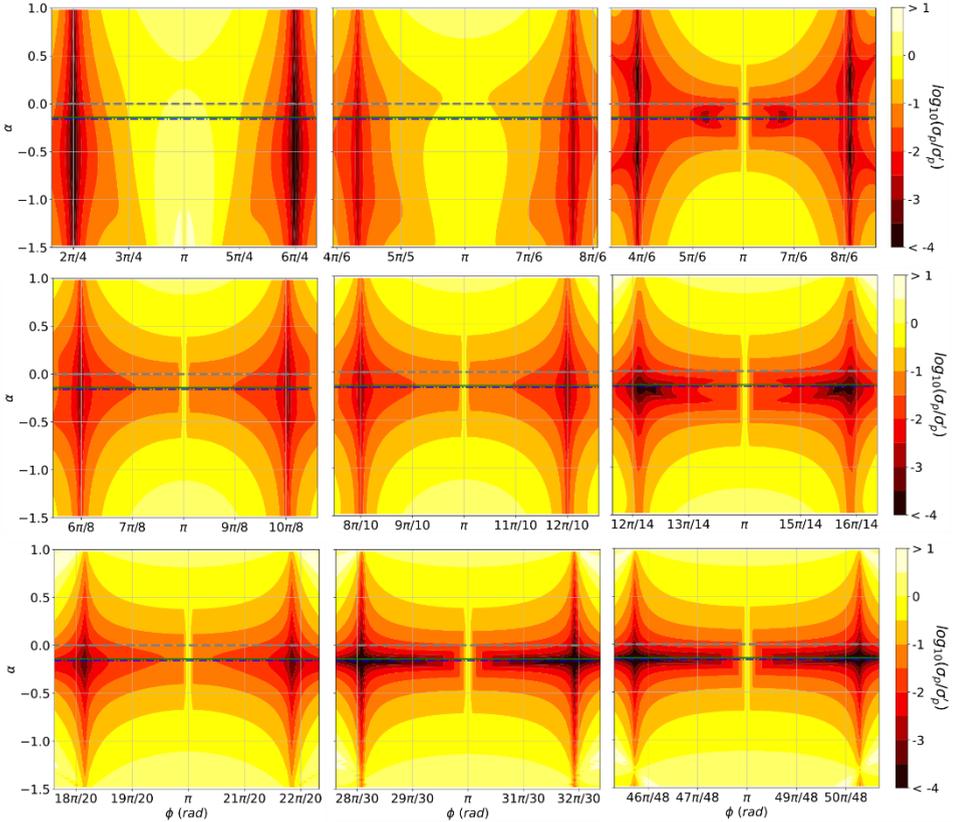

**Fig. E8** (Color online) Relative stability of the quantum random walk search algorithm with walk coin size $m = 2,3,4$ (first row, from left to right), $m = 5,6,7$ (second row), and $m = 8,9,10$ (third row), given by the ratio of the root-mean-square uncertainty with and without the proposed optimization of the coin. The chosen ranges of the parameter $\phi$ correspond to the high probability central peaks of $p(\phi, \zeta = \pi, m)$ (Eq. (7)). The dashed gray, dash-dotted blue, and the solid green lines are obtained from (Eq. (8)) with $\alpha = 0$, $\alpha = -1/(2\pi)$, and $\alpha = \alpha_{ML}$.

# Appendix F    Deep network model and machine learning predictions

Simulations of quantum algorithms with large number of qudits by classical computers are very demanding to the computational resources. This requires a search of alternative means for approaching the task. In this paper predictions of the studied quantum algorithm's parameters for bigger walk coin size, and respectively higher searchable space size, are made with machine learning methods. Here, by Monte Carlo simulations of quantum random walk search algorithm with different size of the coin register, we are able to obtain larger set of independent data that is used to train a machine learning model.

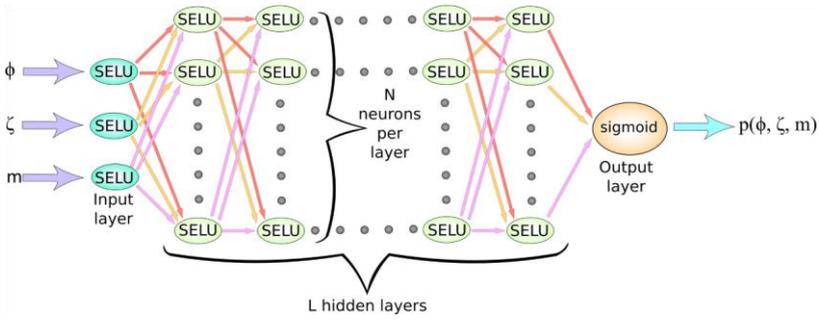

**Fig. F8** (Color online) Scheme of the Deep neural network used for prediction of the probability $p(\varphi, \zeta, m \geq 11)$.

In our previous work [34] a feed-forward deep neural network (DNN) was build using limited number of generated datapoints to train it. The neural network consists of L hidden layers, each with N neurons. The structure of the model used is shown on Fig. F8. In this paper, the data from Monte Carlo simulations for QRWS algorithm with coin size $2 \leq m \leq 10$ is used to augment and fine-tune that model. It is achieved by feeding the DNN with datapoints values of the walk coin parameters $\phi, \zeta, m$, and their labels are the MC results for the probability $p(\phi, \zeta, m)$. Next, the machine learning model is trained for a large number of epochs. Finally, it is used for predictions of the three-parameter function $p(\phi, \zeta, m)$, where $\phi \in [0,2\pi], \zeta \in [0,2\pi], m \geq 2$ (the results for coin size 11 and 16 are plotted on Fig. 3). The values of the parameters $\alpha$ and $\varepsilon$ defined in Sec. 4.1 are extracted by analyzing the above relations. Although our machine learning model gives relatively reliable predictions for number of characteristics of the quantum random walk search algorithm that are explained in the paper, others are not so good. It is well known fact that using DNN for interpolation gives very good results. However, extrapolations away from the training region of the model result in inaccuracies of the predictions. In our case they mainly manifest in small asymmetric widening of the central high probability strip of $p(\phi, \zeta, m)$ (Fig. 3 – middle and right pictures) that leads to some overestimation of the width $\varepsilon$, particularly for higher coin size m, seen on Fig. 6. These deviations are more clearly visible when we focus on the highly nonlinear part of the function $p(\phi, \zeta, m)$ shown on Fig. 2 for $m = 11,16$.